\documentclass[authoryear,preprint,review,12pt]{elsarticle}

\usepackage{amssymb}

\journal{Chaos, Solitons \& Fractals}

\begin{document}

\begin{frontmatter}



\title{Toy nanoindentation model and incipient plasticity}

\author[uc3m,um2]{I. Plans}
 \ead{plans@lcvn.univ-montp2.fr}
\author[ucm]{A. Carpio}
 \ead{carpio@mat.ucm.es}
\author[uc3m]{L. L. Bonilla}
 \ead{bonilla@ing.uc3m.es}

\address[uc3m]{G. Mill\'an Institute for Fluid Dynamics, Nanoscience and Industrial 
Mathematics, Universidad Carlos III de Madrid, 28911 Legan\'es, Spain.\\
}
\address[um2]{Laboratoire des Collo\"ides, Verres et Nanomat\'eriaux, UMR 5587, Universit\'e Montpellier II and CNRS,
34095 Montpellier, France.}
\address[ucm]{Departamento de Matem\'{a}tica Aplicada, Universidad
Complutense de Madrid, 28040 Madrid, Spain}



\begin{abstract}
A toy model of two dimensional nanoindentation in finite crystals is proposed. 
The crystal is described by periodized discrete elasticity whereas the indenter is a rigid strain 
field of triangular shape representing a hard knife-like indenter. Analysis of the model shows 
that there are a number of discontinuities in the load vs penetration depth plot which 
correspond to the creation of dislocation loops. The stress vs depth bifurcation diagram of the 
model reveals multistable stationary solutions that appear as the dislocation-free branch of 
solutions develops turning points for increasing stress. Dynamical simulations show that an
increment of the applied load leads to nucleation of dislocation loops below the nanoindenter 
tip. Such dislocations travel inside the bulk of the crystal and accommodate at a certain depth 
in the sample. In agreement with experiments, hysteresis is observed if the stress is decreased 
after the first dislocation loop is created. Critical stress values for loop creation and their final 
location at equilibrium are calculated.
\end{abstract}


\end{frontmatter}


\section{Introduction}
\label{}

Regularizations of continuum theories typically resolve the difficulties of the latter at small 
scales and often allow precise calculations and descriptions at these scales. This is the case of
the Navier-Stokes equations which regularize the inviscid Euler equations of Fluid Mechanics 
\cite{ll6}, of the lattice regularizations of Quantum Field Theory and phase transitions 
\cite{amit}, etc. Classical elasticity theory is unable to describe the nucleation and motion 
of crystal defects \cite{hul01}, dislocations \cite{hir82}, cracks \cite{slepyan,pla}, 
phase boundaries \cite{zhe08}, or more complex phenomena such as friction 
\cite{rub04,mar04,ger01,kes01}. Lattice regularizations of elasticity can describe the 
structure and motion of nucleated defects and can be analyzed to extract qualitative and 
quantitative information about the phenomena at hand so as to provide a deep physical 
understanding \cite{ger01,pla}. In the case of dislocations, periodized discrete elasticity 
can describe depinning of dislocations at the Peierls stress \cite{CB03}, dislocation cores 
and dislocation interaction \cite{CB}, stable defects corresponding to dislocations in 
graphene membranes and instability of Stone Wales defects \cite{graphene1,graphene2}, and 
homogeneous nucleation of dipoles in a sheared lattice \cite{EPL}. 

Metals usually contain a great number of dislocations whose motion, creation, annihilation 
and interaction are largely responsible for plastic behavior \cite{hir82}. Introducing defects 
in a crystal typically impedes dislocation motion and multiplication thereby strengthening the 
material (strain hardening). At small length scales (such as those intervening in compression
of thin whiskers), dislocations may leave the sample which results in its hardening via 
dislocation starvation \cite{gre06} as observed in compression of nanopillars \cite{sha07}. 
Incipient plasticity occurs when defects are created in a hitherto perfect crystal by different 
means. Nanoindentation experiments are excellent ways to probe incipient plasticity 
\cite{lor03,ase06,rod02,nav08} and so are indentation experiments in colloidal crystals 
\cite{sch06}. In these experiments, the penetration depth inside the crystal is measured as 
the load on the indenter increases, which results in a discontinuous load versus depth diagram. 
The discontinuities in the diagram are thought to indicate dislocation nucleation inside the 
crystal. Different types of calculations have been used to interpret nanoindentation, ranging 
from atomistic simulations to continuum mechanics interpretations or combinations thereof 
\cite{bulatov,kel98,rod02,zhu04}. 

In this paper we present and analyze an atomistic toy model of nanoindentation to show that 
discontinuities in the load vs penetration length diagram do correspond to nucleation of 
dislocation loops \footnote{In a 2D model, dislocation dipoles are pairs of
edge dislocations of opposite Burgers vectors with infinitely long extra half rows of atoms that
go from the dislocation point outwards. Dislocation loops are pairs of edge dislocations with
opposite Burgers vectors sharing a finite segment of extra atoms that join their respective
dislocation points. See Figures 10 and 9 of \cite{CB}.}. We also use this model and its analysis to find different loading-unloading stress vs depth curves, in agreement with experimental observations \cite{lor03}. We have used the AUTO software 
\cite{auto} to find and numerically continue the branches of stationary configurations of our 
model starting from the perfect lattice. Thus we are able to calculate both {\em stable and 
unstable} stationary configurations and the bifurcation diagram (maximum 
strain/dimensionless stress $F$ vs penetration depth $\delta$), which is something that 
cannot be done using Molecular Dynamics (MD) simulations due to neighbor upgrading 
protocols. The bifurcation diagram exhibits multistability of stationary solutions whose 
configurations display an integer number of dislocation loops. Starting from the unstressed 
crystal and as $F$ increases, $\delta$ increases until a turning point\footnote{Points with 
zero or infinite slope where a branch of solutions changes from stable to unstable are turning 
points. See Fig.\ II.1 of \cite{iooss}.} is reached. As $F$ 
surpasses a critical value $F_{c}$, a dislocation loop is nucleated and the profile evolves to 
that of the next stable solution branch in the bifurcation diagram (having a larger $\delta$). 
This gives rise to a discontinuity in the load vs depth diagram for static solutions. Further 
increments of the load give rise to new discontinuities when successive turning points of the 
different stable solution branches are reached. In this simple model, nucleation of dislocation 
loops is a first order phase transition and hysteresis is possible: for $F>F_{c}$, there is an 
abrupt jump of $\delta$ due to dislocation loop nucleation and the unloading curve follows 
the second stable branch which has larger $\delta$ for the same $F$ as in the loading curve. 
For our simple model, the second stable branch ends at a positive value of $F$. Further 
decrease thereof provokes reabsorption of the nucleated dislocation loop at the indenter tip 
and a sudden decrease of $\delta$ to its value on the elastic curve. 

\begin{figure}
\begin{center}
\includegraphics[width=8cm]{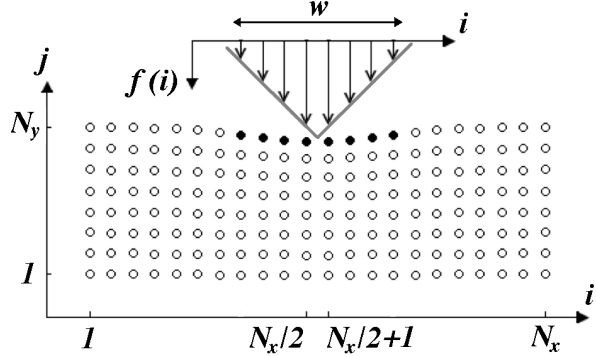}
\caption{Modeling of a nanoindentation experiment. The stress field scales with the 
numerical continuation parameter $F$, such that the maximum downward vertical 
displacement at the central part of the upper surface satisfies $D^-_{2}v_{N_{x}/2,N_{y}}
=- F$.}
\label{fig1}       
\end{center}
\end{figure}

Our toy model is as follows. Consider a 
2D simple cubic lattice with displacement vector measured in units of the lattice constant $l$, 
lattice points labelled by indices $(i,j)$, $i=1,\ldots,N_x$ and $j=1,\ldots,N_y$. The 
nanoindenter at the central region of the upper surface is a hard `knife' 
represented by a fixed strain field which decreases linearly from zero to $-F$ and then 
increases to zero again, as depicted in Fig.~\ref{fig1}. At the other boundaries, the 
displacement vector is zero. The indenter may cause gliding of atom columns in the vertical 
direction and therefore the displacement vector changes more in the vertical direction than in 
the horizontal one. Then a simple displacement vector $(0,v_{i,j})$ captures the qualitative 
features of the physical system. It is convenient to use coordinates $(x,y)$ with $x=i-(N_{x}+
1)/2$ and $y=j-(N_{y}+1)/2$ whose origin is the lattice center. The 
displacements obey the following nondimensional equations: 
\begin{eqnarray}
m {d^2 v_{i,j} \over d t^2} + \alpha {d v_{i,j} 
\over d t} =  v_{i,j+1}-2v_{i,j}+v_{i,j-1} + A \, [g_{a}(v_{i+1,j}-v_{i,j}) + g_{a}(v_{i-1,j}-v_{i,j}) ], \label{sh1}
\end{eqnarray}
where $g_{a}(x)$ is a one-parameter family of periodic functions of $x$ with period 1
\begin{eqnarray}
g_a(x)= {2a\over \pi}\left\{\begin{array}{ll}
\sin\left({\pi x\over 2 a}\right), & -a \leq x \leq a,\\
\sin\left({\pi(x-1/2)\over 2a-1}\right), & a \leq x \leq 1-a,\\
\end{array}\right.
\label{sh3}
\end{eqnarray}
and $0< a< 1/2$, such that $g_{a}(x)\sim x$ as $x\to 0$ (see below for motivation). The 
boundary conditions are 
\begin{eqnarray}
&&  D^-_{2} v_{i,N_{y}}\equiv v_{i,N_{y}}-v_{i,N_{y}-1}= -F\, f(i), \label{bc1}\\
&& v_{i,j} =  0, \quad\mbox{if either } j=1,\, i= 1, \mbox{ or } i=N_{x}. \label{bc2}
\end{eqnarray}
For a symmetric and centered sharp indenter as depicted in Fig.~\ref{fig1}
with an even number of atoms, $w$, and for even $N_{x}$, we have
\begin{equation} \label{ap:a:sine}
f(i) =\frac{2}{w}\times \left\{\begin{array}{cc} 
i-\frac{N_x - w}{2}, &\frac{N_x - w}{2}+1 \le i \le \frac{N_x}{2}, \\
\frac{N_x + w}{2}+1-i, & \frac{N_x}{2}+1 \le i \le \frac{N_x + w}{2},\\
0, & \mbox{otherwise.}
\end{array} \right.
\end{equation}
This indenter has a rectangular cross section $S=Lwl$, $wl\ll L$, where $l$ is the lattice
constant and $L$ is the length of the knife. A more realistic modeling of the knife would require
considering horizontal displacements which we are ignoring in our model. Note that $f(N_{x}
/2)=f(1+N_{x}/2)=1$ and that $\sum_{i} f(i)=1+w/2$. In Eq.\ (\ref{sh1}), 
$A=C_{44}/C_{11}$ provided we consider cubic crystals with elastic constants $C_{11}$, 
$C_{12}$, $C_{44}$. The vertical component of the stress tensor $\sigma_{22}$ is simply 
$C_{11}\, D^-_{2}v_{i,j}$ in our model, and therefore the strain at the surface given by 
Eq.\ (\ref{bc1}) is also the nondimensional applied stress $\sigma_{22}/C_{11}$. 
We also have $\sigma_{21}/C_{11}= A\, g_{a}(D_{1}^-v_{i,j})/2$. The 
relation between the stress at the surface and an applied load $P$ is 
\begin{eqnarray}
P= \frac{S}{w}\sum_{i=2}^{N_{x}}\sin\varphi(i)\,\sigma_{21}(i,N_{y})
- \frac{S}{w}\sum_{i=2}^{N_{x}}\cos\varphi(i)\,\sigma_{22}(i,N_{y}), \nonumber
\end{eqnarray}
where $\varphi(i)$ is the angle between the normal to the indenter surface and the $y$ axis, 
with $\tan\varphi=D^-_{1} v_{i,N_{y}}$. We obtain 
\begin{eqnarray}
\Lambda\equiv\frac{P}{C_{11}S}= \frac{F}{w}\,\sum_{i=2}^{N_{x}}\cos
\varphi(i)\, f(i) 
+ \frac{A}{2w}\sum_{i=2}^{N_{x}}\sin\varphi(i)\, g_{a}(D^-_{1}
v_{i,N_{y}}),   \label{load}
\end{eqnarray}
which relates the nondimensional load $\Lambda$ to the stress $F$. If we select a 
nondimensional time scale $C_{11}t/(\rho l^2\gamma)\to t$, then $\alpha=1$ and $m=
C_{11}/(\rho l^2\gamma^2)$, where $\gamma$ is a friction coefficient with units of 
frequency, $\rho$ is the mass density and $v_{i,j}$ is measured in units of $l$. 
With this choice of scales, we can consider the overdamped case with $m=0$. On the other 
hand, if we select a nondimensional time scale $C_{11}^{1/2}t/(l\rho^{1/2})\to t$, then $m
=1$ and $\alpha=l\gamma\sqrt{\rho/C_{11}}$. With this second choice of scales, we can 
consider the conservative case with $\alpha=0$. 

\begin{figure}
\begin{center}
\includegraphics[width=8cm]{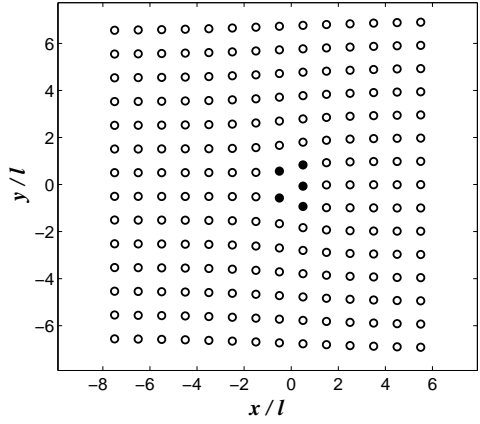}
\caption{An edge dislocation parallel to the $z$ axis with Burgers vector $(0,-1)$. }
\label{fig0}     
\end{center}
\end{figure}

If $g_{a}(x)=x$ in Eq.\ (\ref{sh1}), we obtain discretized scalar linear elasticity. Why do 
we have to use the periodic functions (\ref{sh3})?  To allow atoms change neighbors 
during dislocation motion. The fact that the functions (\ref{sh3}) are periodic allows the 
atoms to change neighbors while the computational grid remains unchanged (and so the
 dynamical system (\ref{sh1}) whose linear stability will be analyzed).

 Let us explain this idea. Assume that we have an extra half row of atoms 
in the square lattice that is parallel to the positive $x$ direction (edge dislocation with Burgers 
vector $(0,-1)$), as depicted in Fig.~\ref{fig0}. If these atoms move one step downwards 
and occupy the equilibrium positions of their neighboring atoms, the latter are displaced 
downwards and become the extra half row representing the edge dislocation that has therefore 
moved one step downwards. If these new atoms move one step downwards in the same way, a 
new half row replaces them and the dislocation has moved another step. During this glide 
motion, the leftmost atom in the half row changes neighbors horizontally and only once 
whereas all the other atoms in the half row continue having the same neighbors. 
Even though the dislocation can glide for a long distance, single atoms move very little and 
only a very small fraction of these atoms change neighbors. 
Instead of updating neighbors during glide motion, we can construct a periodized discrete 
elasticity model by using a nonlinear periodic function $g_a$ in (\ref{sh1}), with period 
equal to the lattice space and $g_a(x)\sim x$ as $x\to 0$. A simple example is (\ref{sh3}). 
This model changes neighbors without updating, thereby allowing
atomic half rows to glide in the y-direction. 

As the applied stress becomes sufficiently large, there appear edge dislocation loops whose 
Burgers vectors are directed along the $y$ axis. Gliding along other directions is not possible 
in this model: we would need a two-component displacement vector and a periodic function 
of discrete differences along the $x$ axis \cite{CB}. The parameter $a$ controls the 
asymmetry of $g_a$: 
$a$ increases, the interval over which $g'_{a}(x)>0$ increases at the expense of the interval 
over which the slope of $g_{a}$ is negative. As $a$ increases, the (dimensionless) Peierls stress $\sigma_p$ needed for a 
dislocation to start moving increases whereas the size of the dislocation core and its mobility 
both decrease \cite{CB}\footnote{Note that the parameter $\alpha$ used in \cite{CB} 
corresponds to $-a+1/2$ in (\ref{sh3}) and therefore the Peierls stress in Figure 2 of 
\cite{CB} decreases as $\alpha$ increases.}. The value of $a$ can be selected so that the 
Peierls stress calculated from (1) fits values measured in experiments or calculated using MD. 
Namely, for a $16\times 30$ lattice with $A=0.2258$ (corresponding to gold) we get 
$\sigma_p = 0.004$ for $a=0.2$ compared to $\sigma_p = 0.03$ for $a=0.4$. 

In the symmetric case $a=1/4$, (\ref{sh1}) and (\ref{sh3}) are the governing equations
of the interacting atomic chains model \cite{lkk}.

\section{Methodology and Results}
We consider parameters $A=0.2258$ (gold), $a=0.2$, for a $22 \times 42$ 
lattice and a four atom indenter, $w=4$. Similar results were obtained when the same indenter 
is placed in the center of the upper surface of other lattices with even $N_{x}$ and $N_{y}$. 
Asymptotically stable stationary solutions of the model equations (\ref{sh1}) with boundary 
conditions (\ref{bc1}) - (\ref{bc2}) are obtained by numerically solving the equations with 
appropriate initial conditions and overdamped dynamics, $m=0$, $\alpha=1$. For conservative
dynamics, $m=1$, $\alpha=0$, the same stationary solutions are stable but not asymptotically
stable. Unstable solutions are unstable no matter which dynamics is considered. 

\begin{figure}
\begin{center}
\includegraphics[width=8.2cm]{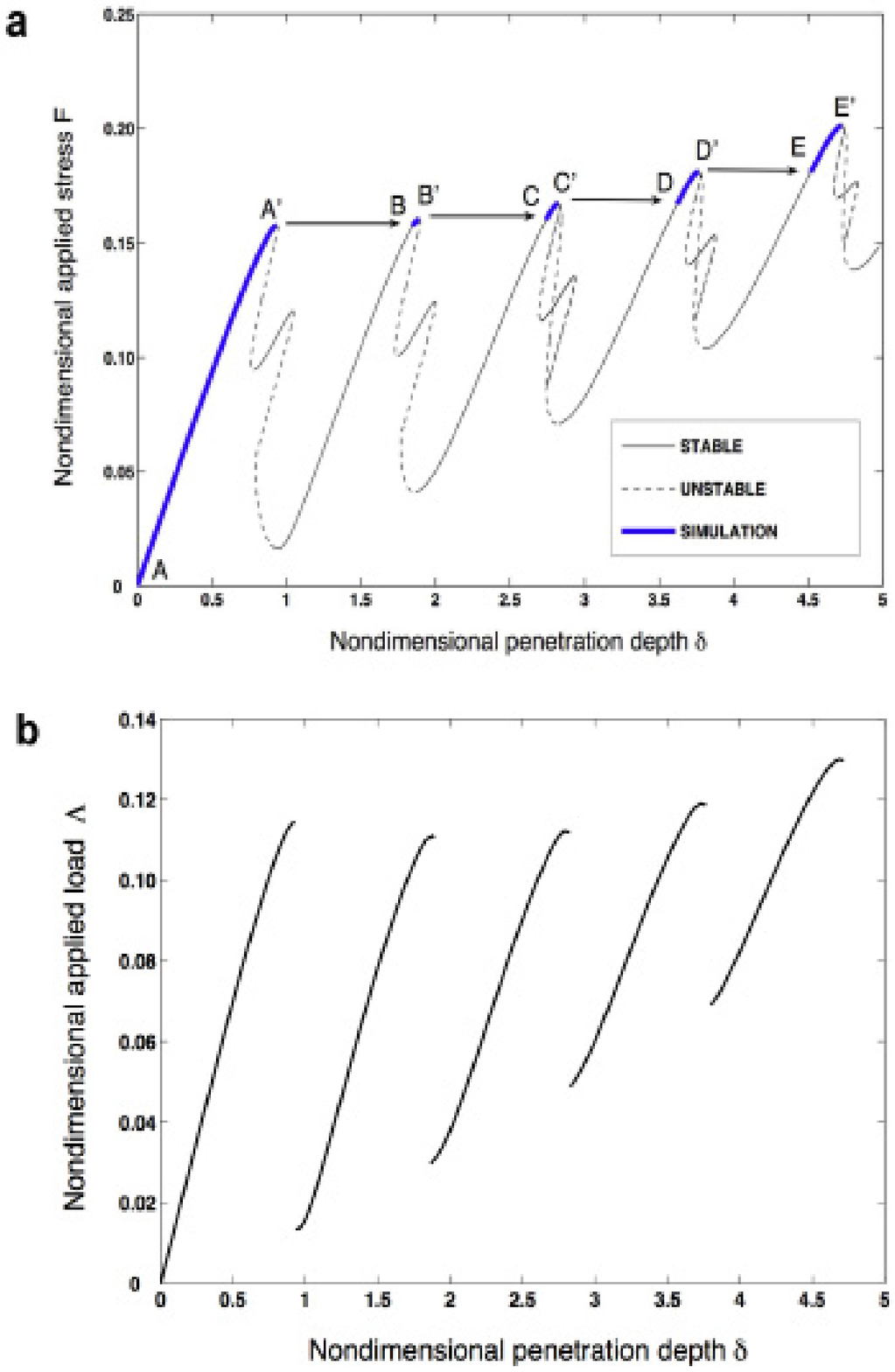}
\caption{(a) Maximum vertical nondimensional stress $F$ \textit{vs} nondimensional
indenter penetration depth $\delta$ obtained by numerical continuation of the stationary 
solution at $F=\delta=0$ for a 16x42 lattice, with $A=0.2258$ (gold), $a = 0.2$. By 
increasing adiabatically $F$ from zero, we sweep this 
bifurcation diagram following the path $A\to A'\to B\to B'\to C\to C'\to D\to D'\to 
E\to E'$. (b) Similar diagram of nondimensional applied load \textit{vs} nondimensional 
penetration depth. Only the long stable portions of the bifurcation diagram are shown.}
\label{fig2}       
\end{center}
\end{figure}

At zero applied load the perfect lattice is found. As the dimensionless applied load $F$ 
increases from zero, we can monitor the penetration depth of the indenter, 
\begin{eqnarray}
\delta= -\frac{1}{w}\,\sum_{i=1+(N_{x}-w)/2}^{(N_{x}+w)/2} v_{i,N_y-1}, 
\label{depth}
\end{eqnarray}
and obtain the relation between $F$ and $\delta$. We have used the AUTO software 
\cite{auto} to find and numerically continue the branches of stationary solutions of our 
model starting from the perfect lattice with $F=0$, $\delta=0$. The results are represented 
in Fig. \ref{fig2}a, whereas Fig.\ \ref{fig2}b shows the corresponding load vs
penetration depth diagram (in dimensionless units). The bifurcation diagram shows a 
connected branch of stationary solutions which begins at the origin (marked with $A$ in Fig.\ 
\ref{fig2}a). Each point of the long stable portions of this branch corresponds to a single 
configuration of the lattice. There are many turning points (local maxima or minima, some 
of them indistinguishable at the scale of the plot) connecting stable and unstable regions. The 
main turning points point out to the occurrence of different nucleation events, which we 
shall now describe. 

\begin{figure}
\begin{center}
\includegraphics[width=8cm]{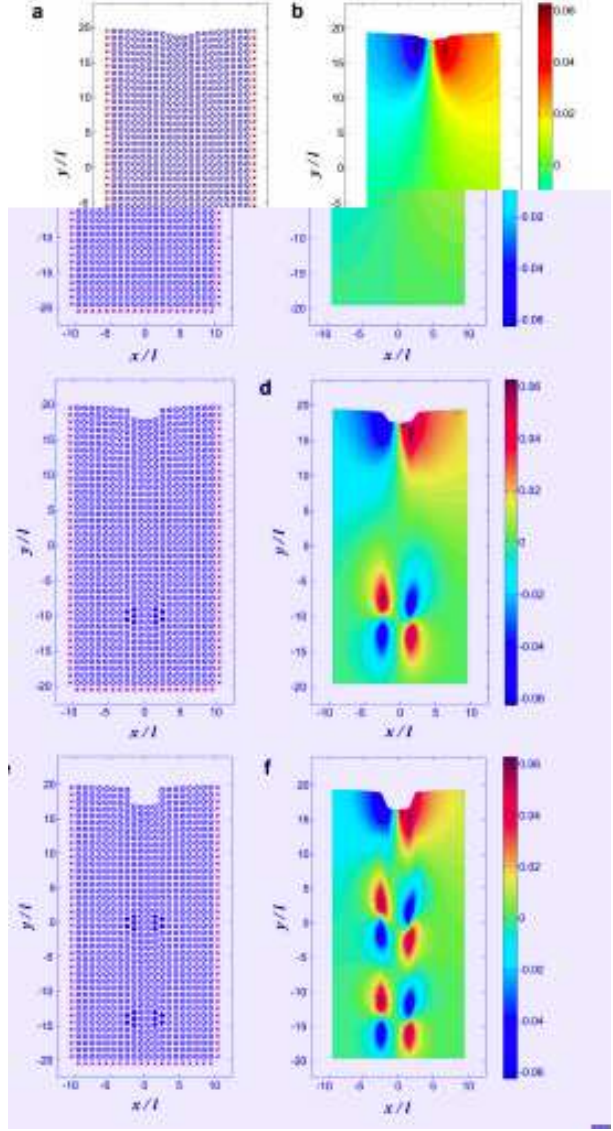}
\caption{Stable configurations at the turning points (a) $A'$, (c) $B'$ and (e) $C'$ of the
bifurcation diagram in Fig.~\ref{fig2}. Panels (b), (d) and (f) visualize the component 
$e_{1,2}=g_a(v_{i+1,j}-v_{i,j})/2$ of the strain tensor corresponding to $A'$, $B'$ and 
$C'$, respectively.}
\label{fig3}       
\end{center}
\end{figure}

The longest stable portions of the stationary branch end at turning points $A'$, $B'$, $C'$, $D'$ 
and $E'$ in increasing order of $\delta$ (solid lines in Fig. \ref{fig2}a). The corresponding 
configurations have 0, 1, 2, 3 and 4 dislocation loops, respectively, as shown in 
Figs.~\ref{fig3}a, c and e. We may call $AA'$ the dislocation-free {\em elastic branch}, 
whereas the other long stable portions of Fig.\ \ref{fig2}a signal the beginning of 
plasticity due to loop formation. The component $e_{1,2}=g_a(v_{i+1,j}-v_{i,j})/2$ of the 
strain field (proportional to the dimensionless shear stress) provides a good visualization of 
the dislocation loops as Figures \ref{fig3}b, d and f show. Each loop comprises two 
dislocations whose cores are located at points with $i=1+(N_{x}-w)/2$ and $i=(N_{x}+w)/2$ 
below the indenter. Using the lattice constant as unit of length, the first dislocation is at the 
column $i=1+(N_{x}-w)/2$ and it has Burgers vector $(0,-1)$, whereas the second dislocation 
is located at the column $i=(N_{x}+w)/2$ and it has Burgers vector $(0,1)$. 

Note that after some of its turning points the solution branch bends over itself, and connects back to earlier limit points ($C'$, $D'$, $E'$ in Fig. \ref{fig2}a) by means of unstable portions. To each point of the short stable portions found after these limit points, there correspond two lattice configurations, which are symmetrical with respect to $x=0$. They contain one extra edge dislocation in addition to the dislocation loops that the configuration corresponding to the 
preceding long stable portion with smaller $\delta$ may contain. In one configuration, the 
dislocation point of the extra dislocation with Burgers vector $(0,1)$ is located at $i=(N_{x}+
w)/2$, at some distance below the indenter (as occurs in the short stable portions after $A'$ and $B'$). The other configuration has a dislocation point on 
the same row, with opposite Burgers vector $(0,-1)$, which is located at $i=1+(N_{x}-w)/2$. 
As it will be explained later, these short stable portions are not reached by any of our dynamical experiments. However, if the configurations corresponding to unstable branches are depicted, nucleation or 
disappearance of the extra edge dislocation is observed as we run AUTO along the unstable 
branch towards the turning point where it joins a stable branch. 

\section{Adiabatic sweeping of the bifurcation diagram}
By increasing adiabatically $F$ from $F=0$ (up-sweep), the stable portions $AA'$, $BB'$, 
$CC'$, $DD'$, $EE'$ of the solution branch in the bifurcation diagram of Fig.~\ref{fig2}a 
are successively swept. $A'$ marks the critical value $F_c$ above which dislocation loops are 
nucleated. $B'$, $C'$, $D'$ and $E'$ are the last points found in the corresponding stable 
portion of the solution branch before a new dislocation loop is nucleated and a jump to 
another stable portion occurs during adiabatic up-sweep. At these points, with stresses 
$F_{A'}$, $F_{B'}$, $F_{C'}$, $F_{D'}$ and so on, $\delta$ jumps to the corresponding 
value in the next branch, as indicated by the arrows in Fig.\ \ref{fig2}a, and a new 
dislocation loop is formed. What is observed during slow up-sweep by numerically solving 
(\ref{sh1})? After $F$ surpasses $F_{A'}$ in Fig.\ \ref{fig2}a with configuration as in
Fig.\ \ref{fig3}a, a dislocation loop is nucleated {\em immediately below the indenter tip} and 
it glides downwards until it reaches its stable position inside the sample at $y/l = -10$ 
(configuration $B$). The jump $A'\rightarrow B$ in Fig.\ \ref{fig2}a has taken place. 
By adiabatically increasing $F$, we may reach $B'$ (cf.\ Fig.\ \ref{fig3}c and d), 
which is almost identical to $B$. After $F$ surpasses $F_{B'}$, a new dislocation loop is 
nucleated at the indenter tip, and now both loops glide downwards until they reach their 
respective stable positions $y/l = 0, -14$ in configuration $C$. Further increase of $F$ leads
to configuration $C'$ (see Fig.\ \ref{fig3}e). Additional up-sweep repeats this pattern: old 
dislocation loops glide downwards and new ones appear beside the indenter tip.

If we start relaxing adiabatically the load on the nanoindenter after $F$ has reached $F_{B}$, 
then the displacement $\delta$ does not go back to the value $\delta_{A'}$. Instead, its 
value decreases following the stable branch containing the point $(\delta_{B},F_{B})$  in 
Fig.\ \ref{fig2}a. As $F$ decreases, the loop created in the middle of the sample glides 
upwards towards the indenter until the turning point with lowest value of $F$ is reached. At 
this stable configuration, the loop is located at $y/l = 5$. An additional decrease of the load 
results in a downward jump to the elastic solution branch. During this dynamical process, the 
dislocation loop glides upwards and disappears as it reaches the indenter (in 3D loops may be 
anchored due to cross-slip or to preexisting defects \cite{hul01}). In a standard model of 
plastic behavior, we would expect that the second stable branch of stationary solutions 
extend to $F=0$, $\delta>0$ without losing the dislocation loop.

There are other stable portions of the stationary solution branch in Fig.~\ref{fig2}a that 
cannot be reached by up- or down-sweeping the bifurcation diagram: the short portions with 
intermediate values of $\delta$ between those of two long stable portions.

\begin{figure}
\begin{center}
\includegraphics[width=8cm]{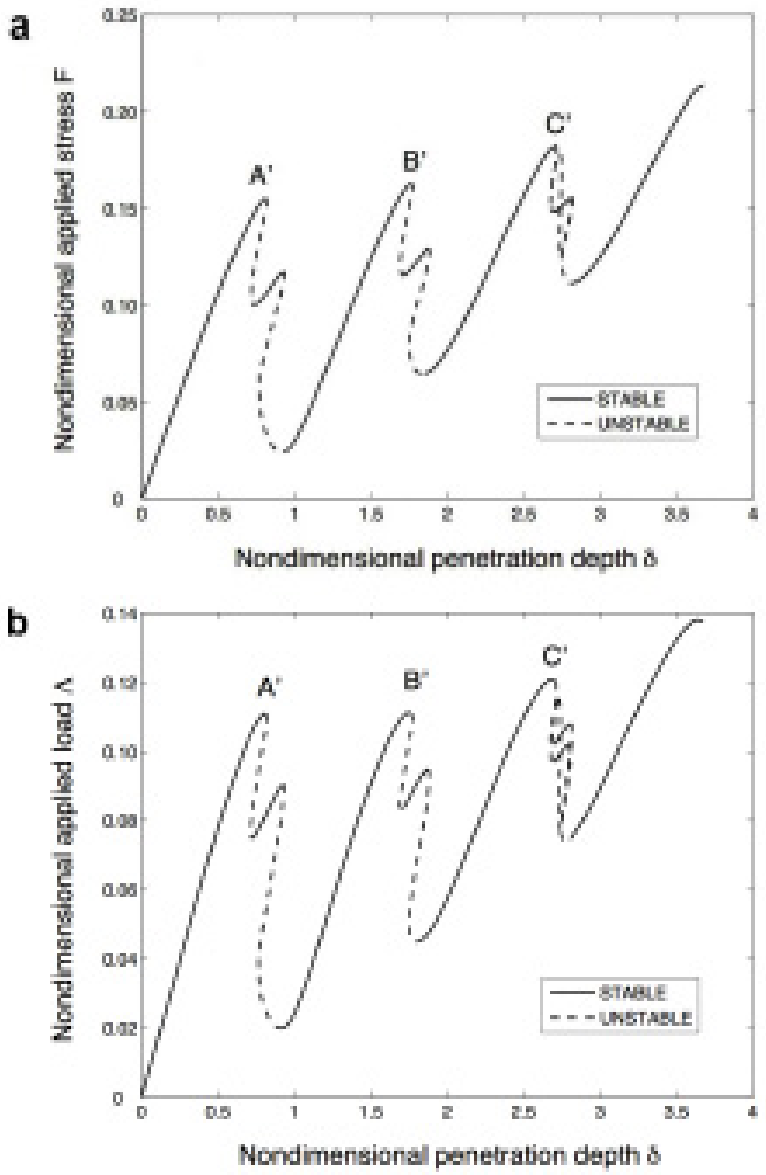}
\caption{Same as in Fig.\ 3 for a 16x30 lattice.}
\label{fig5}       
\end{center}
\end{figure}

\section{Results for other parameter values} 
For the same values of $a$ and $A$, we have changed the position of the indenter, made it
wider, changed the size of the lattice or considered odd $N_{x}$. In all these cases, the 
$F$ (or $\Lambda$) vs $\delta$ diagram is similar to that in Fig.~\ref{fig2}, as shown in 
Fig.~\ref{fig5} for a lattice of different size. The unstable 
portions of the bifurcation diagram change although not the long stable portions, and 
therefore the responses to adiabatic up- and down-sweeping the diagram are the same. For 
parameter values corresponding to the point $B'$, the dislocation loop is located at a height
roughly two thirds of $N_{y}$ \footnote{For $N_{x}=16$, $\delta_{B'}/N_{y}= 0.625$, 
0.661 and 0.668 for $N_{y}=30$, 42 and 58, respectively.}. 
However there are some differences worth noticing. For a $22\times 42$ lattice, increasing 
the indenter width from $w=4$ to $w=10$ increases only slightly $F_{A'}$, from 0.1575 to 
0.1598, but now the cavities created by the indenter are not rectangular as in Fig.~\ref{fig3}. 
Instead, they exhibit steps which made them triangular in shape. Similar results are found
for odd $N_{x}$ when the maximum $f(i)$ is reached at only one atom, $i=[N_{x}/2]$.
For a $17\times 30$ lattice with $w=5$, the bifurcation diagrams are similar and triangular 
cavities are observed. A shorter indenter with $w=4$ placed on atoms 7, 8 , 9 and 10 of a 
$17\times 30$ lattice again gives rise to bifurcation diagram similar to Fig.~\ref{fig5} but 
now some of the short intermediate stable portions having one extra unpaired edge
dislocation can be distinguished in the diagram because their corresponding configurations are
no longer symmetric. Actually, it is not always after the third turning point ($C'$ in Figs.~\ref{fig2} and~\ref{fig5}) that these short intermediate stable portions correspond to two solutions. We find that for a $22 \times 22$ lattice it occurs after the second turning point, $B'$.
 If we use $A=0.448$ (corresponding to copper), for a $16
\times 30$ lattice the results are qualitatively similar to those so far presented except that: (i) 
the point $A'$ has a larger $F=0.2191$ and a smaller $\delta=0.7958$ (compared to $F=
0.1553$ and $\delta=0.8128$ for gold in Fig.~\ref{fig5}); and (ii) close to the minima of 
the bifurcation diagram, the branch of stationary solutions is wiggly with several turning points 
separating alternating stable and unstable solutions. These wiggly portions are also present in 
Fig.~\ref{fig5} but they are not visible at the scale chosen in the figure. 

As mentioned before, augmenting $a$ in Eq.\ (\ref{sh3}) increases $F_{A'}$ and the Peierls stress and reduces 
the size of defect cores. For a $16\times 30$ lattice with $w=4$, $A=0.2258$ (gold) and $a=
0.4$, we get $F_{A'}= 0.2558$, $\delta_{A'}= 1.2283$ (compared to $F=0.1553$ and 
$\delta=0.8128$ for $a=0.2$) and it is quite hard to move dislocations (the dimensionless 
Peierls stress is $\sigma_p=0.03$, compared to $\sigma_p=0.004$ for $a=0.2$). The bifurcation 
diagram changes in that the number of turning points increases enormously. However, 
the stable portions of the diagram have the same configurations and interpretations as those in 
Figs.~\ref{fig2} and \ref{fig5}. Then up-sweeping the diagram produces the same 
sequence of discontinuities due to formation of dislocation loops.

\section{Conclusion}\label{conclusion}
We have presented a toy model of 2D nanoindentation and incipient plasticity in a 
dislocation-free crystal based on periodized discrete elasticity. Analysis and numerical 
simulation of the model confirm that the discontinuities in the diagram of stress (or load) vs 
penetration depth are associated with the nucleation of dislocation loops (if stress is 
adiabatically increased) or with their absorption in contact with it (if stress is 
adiabatically decreased). Our simulations and analysis of the bifurcation diagram of solutions 
show that hysteresis occurs, leading to unloading curves which are different to the loading 
ones, in agreement with experimental observations. The analysis of the results obtained with 
AUTO yields the critical stress for dislocation nucleation, the type thereof and the depth at 
which dislocation loops are accommodated in their equilibrium positions.

\section*{Acknowledgments}
We thank O. Rodr\'\i guez de la Fuente and J.M. Rojo for fruitful discussions and useful
comments. This work 
has been supported by the Spanish Ministry of Science and Innovation grants 
FIS2008-04921-C02-01 (LLB and IP) and FIS2008-04921-C02-02 (AC), by the 
Autonomous Region of Madrid under grant S-0505/ENE/0229 (COMLIMAMS) (LLB and
IP) and by CM/UCM 910143 (AC).





\begin{thebibliography}{00}

\bibitem{ll6}
Landau LD and Lifshitz EM. Fluid Mechanics (2nd ed.): Pergamon P., New 
York; 1987.

\bibitem{amit}
Amit DJ and Mart\'\i n Mayor V. Field Theory, the Renormalization Group 
and Critical Phenomena (3rd. rev. ed.): World Sci. Singapore; 2005.  

\bibitem{hul01} 
Hull D and Bacon DJ. Introduction to Dislocations (4th ed.): Butterworth-Heinemann, Oxford UK; 2001.

\bibitem{hir82} 
Hirth JP, Lothe J. Theory of Dislocations (2nd ed.): John Wiley and Sons, New 
York; 1982.

\bibitem{slepyan}
Slepyan LI. Models and Phenomena in Fracture Mechanics: Springer, Berlin 2002.

\bibitem{pla}
Pla O, Guinea F, Louis E, Ghaisas SV and Sander LM. Straight cracks in dynamic brittle fracture. Phys Rev B 2000;61:11472-11486.

\bibitem{zhe08} 
Zhen Y and Vainchtein A. Dynamics of steps along a martensitic phase boundary I: Semi-analytical solution. J Mech Phys Solids 2008;56:496-520.

\bibitem{rub04}
Rubinstein SM, Cohen G and Fineberg J. Detachment fronts and the onset of dynamic friction. Nature 2004;430(7003):1005-1009.

 \bibitem{mar04}
Marder M. Friction - Terms of detachment. Nature Materials 2004;3(9):583-584.

\bibitem{ger01} Gerde E and Marder M. Friction and fracture. Nature 2001;413(6853),285-288.

 \bibitem{kes01} Kessler DA. Surface physics - A new crack at friction. Nature 2001;413(6853),260-261.
 
 \bibitem{CB03}
Carpio A and Bonilla LL. Edge dislocations in crystal structures considered as traveling waves in discrete models. Phys Rev Lett 2003;90:135502.

\bibitem{CB}
Carpio A and Bonilla LL. Discrete models of dislocations and their motion in cubic crystals. Phys Rev B 2005;71(13):134105.

\bibitem{graphene1}
Carpio A, Bonilla LL, de Juan F and Vozmediano MAH. Dislocations in graphene. New J Phys 2008;10:053021.

\bibitem{graphene2}
Carpio A, Bonilla LL. Periodized discrete elasticity models for defects in graphene. Phys Rev B 2008;78(8):085406.

\bibitem{EPL}
Plans I, Carpio A and Bonilla LL. Homogeneous nucleation of dislocations as bifurcations in a periodized discrete elasticity model. Europhys Lett 2008;81(3):36001.

\bibitem{gre06}
Greer JR and Nix WD. Nanoscale gold pillars strengthened through dislocation starvation. Phys Rev B 2006;73(24):245410.

\bibitem{sha07}
Shan ZW, Mishra RK, Asif SAS, Warren OL and Minor AM. 
Mechanical annealing and source-limited deformation in submicrometre-diameter Ni crystals. Nature Materials 2008;7(2):115-119.

\bibitem{lor03}
Lorenz D, Zeckzer A, Hilpert U, Grau P, Johansen H and Leipner HS. Pop-in effect as homogeneous nucleation of dislocations during nanoindentation. Phys Rev B 2003;67(17):172101.

\bibitem{ase06}
Asenjo A, Jaafar M, Carrasco E and Rojo JM. Dislocation mechanisms in the first stage of plasticity of nanoindented Au(111) surfaces. Phys Rev B 2006;73(7):075431.

\bibitem{rod02}
de la Fuente OR, Zimmerman JA, Gonz\'alez MA, de la Figuera J,
Hamilton JC, Pai WW and Rojo JM. Dislocation emission around nanoindentations on a (001) fcc metal surface studied by scanning tunneling microscopy and atomistic simulations. Phys Rev Lett 2002;88(3):036101.

\bibitem{nav08}
Navarro V, de la Fuente OR, Mascaraque A and Rojo JM. Uncommon Dislocation Processes at the Incipient Plasticity of Stepped Gold Surfaces. Phys Rev Lett 2008;100(10):105504.

\bibitem{sch06} 
Schall P, Cohen I, Weitz DA and Spaepen F. Visualizing dislocation nucleation by indenting colloidal crystals. Nature 2006;440(7082):319-323.

\bibitem{bulatov}
Bulatov VV and Cai W. Computer simulations of 
dislocations: Oxford University Press, Oxford, UK; 2006.

\bibitem{kel98}
Kelchner CL, Plimpton SJ and Hamilton JC. Dislocation nucleation and defect structure during surface indentation. Phys Rev B 1998;58(17):11085-11088.

\bibitem{zhu04}
Zhu T, Li J, Van Vliet KJ, Ogata S, Yip S and Suresh S. Predictive modeling of nanoindentation-induced homogeneous dislocation nucleation in copper. J Mech
Phys Solids 2004;52(3): 691-724.

\bibitem{auto}
Doedel EJ, Paffenroth RC, Champneys AR, Fairgrieve TF, Kuznetsov YA, Oldeman BE, Sandstede B and Wang X. AUTO$2000$: Continuation and bifurcation 
software for ordinary differential equations (with HomCont). Technical Report, Concordia 
University;  2002. 
{\tt https://sourceforge.net/projects/auto2000/},
{\tt http://indy.cs.concordia.ca/auto/}

\bibitem{iooss}
Iooss G and Joseph DD. Elementary Stability and Bifurcation Theory: Springer, New 
York; 1980. 

\bibitem{lkk}
Landau AI. Application of a model of interacting atomic chains for the description of edge dislocations. Phys Status Solidi B 1994;183(2),407-417.



\end{thebibliography}
\end{document}